\documentclass[conference]{IEEEtran}

\usepackage{xspace}
\usepackage{graphicx}
\usepackage{amsmath}
\usepackage{booktabs}
\usepackage{xcolor}

\usepackage[utf8]{inputenc}
\usepackage{subcaption}
\usepackage{siunitx}

\newcommand{\para }[1]{\medskip \noindent  {\bf #1}}
\newcommand{\1}{{\em (i)}}
\newcommand{\2}{{\em (ii)}}
\newcommand{\3}{{\em (iii)}}
\newcommand{\4}{{\em (iv)}}
\newcommand{\5}{{\em (v)}}
\newcommand{\6}{{\em (vi)}}

\newcommand{\sysname}{\textsc{Erhard-Rng}\xspace}

\newcommand{\thetitle}{\sysname{}: A Random Number Generator Built from Repurposed Hardware in Embedded Systems}

\newcommand{\ra}[1]{\renewcommand{\arraystretch}{#1}}

\begin{document}

\title{\thetitle}

\author{\IEEEauthorblockN{Jacob Grycel and Robert J. Walls}
\IEEEauthorblockA{Department of Computer Science\\
Worcester Polytechnic Institute\\
Worcester, MA\\
Email: {jtgrycel, rjwalls}@wpi.edu}}
\maketitle

\begin{abstract} 

Quality randomness is fundamental to cryptographic operations but on embedded
systems good sources are (seemingly) hard to find.  Rather than use expensive
custom hardware, our \sysname Pseudo-Random Number Generator (PRNG) utilizes
entropy sources  that are already common in a range of low-cost embedded
platforms. We empirically evaluate the entropy provided by three sources---SRAM
startup state, oscillator jitter, and device temperature---and integrate those
sources into a full Pseudo-Random Number Generator implementation based on
Fortuna~\cite{fortuna}. Our system addresses a number of fundamental challenges
affecting random number generation on embedded systems. For instance, we propose
SRAM startup state as a means to efficiently generate the initial seed---even
for systems that do not have writeable storage. Further, the system's use of
oscillator jitter allows for the continuous collection of entropy-generating
events---even for systems that do not have the user-generated events that are
commonly used in general-purpose systems for entropy, e.g., key presses or
network events.

\end{abstract} 
\IEEEpeerreviewmaketitle

\section{Introduction}
\label{sec:intro}

Modern security technologies depend on strong random numbers for creating
encryption keys, signatures, and nonces for sensitive data.  The strength of
these numbers is dependent on a reliable source of entropy and a secure
Pseudo-Random Number Generator (PRNG). On general purpose systems such as
Linux, the entropy is usually gathered from the timing of unpredictable events,
e.g., user key strokes, mouse movements, and disk events. Further, initial
seeding of the generator is usually done by reading in a seed file or by
delaying random number generation until enough entropy can be collected.
However, entropy sources used by general purpose systems are poorly suited to
the embedded environment. For example, key stroke information cannot be
collected by a system that does not have a keyboard attached. Similarly, seed
files cannot be used when the system does not have a disk.    
 
This work examines the feasibility of implementing a PRNG fed by high-quality
entropy sources in an embedded environment with limited hardware and software
resources. Previous studies proposed using specialized hardware in either an
FPGA or an ASIC to gather entropy~\cite{osc-gen, osc-gen-cmos}; However, such
hardware can add significant production cost to embedded systems.  In contrast,
the system we present, \sysname, utilizes existing internal hardware that is
commonly found in a wide variety of inexpensive microcontrollers. In
particular, our implementation of \sysname targets the Texas Instruments MSP430
(MSP430F5529) low-power microcontroller using only an external \SI{4}{\MHz}
crystal oscillator and power supply. We chose this platform due to its
numerous programmable peripherals, its commonality in embedded systems
development, and its affordability. Many microcontrollers and Systems On a Chip
(SoC) include similar hardware that allows our PRNG to be implemented,
including chips from Atmel, Xilinx, Cypress Semiconductor, and other chips from
Texas Instruments.

\sysname is based on three key insights. First, we can address the challenge of
collecting entropy at runtime by sampling the jitter of the
low-power/low-precision oscillator. Coupled with measurements of the internal
temperature sensor we can effectively re-seed our PRNG with minimal overhead.
Second, we can solve the problem of initial seeding by leveraging 
randomness inherent in the startup state of RAM~\cite{holcomb}. Finally, we
also propose the use of a Cyclic Redundancy Check (CRC) as a mixing and
collapsing function for transforming the RAM startup state into the initial
seed, overcoming the processing limitations created by a small memory space.

Our system, based on the Fortuna PRNG~\cite{fortuna}, was designed and
implemented on the MSP430. We empirically evaluate the amount of entropy
generated by each of our sources and the overall quality of the random numbers
generated by the PRNG. Additionally, we analyze the effect of different
operating environments on the entropy sources; As systems operate at different
supply voltages and environment temperatures, it is important to understand the
extent to which the performance of the PRNG is dependent on particular physical
parameters, e.g., external temperature.
 \section{Design and Implementation of \sysname}
\label{sec:prng}

PRNGs rely on a source of true randomness, entropy, that often provides an
initial seed from which a random stream of data is produced. Depending on the
generator design, entropy sources may also be used to periodically re-seed the
PRNG, extending the number of outputs that can be produced before the generator
is liable to repeat sequences or become predictable. One such generator that
frequently uses entropy sources to re-seed is Fortuna, a block cipher-based PRNG
designed by Niels Ferguson and Bruce Schneier~\cite{fortuna}. Fortunta forms the
foundation of \sysname.

\sysname, like Fortuna, is divided into two components: the \emph{generator}
and the \emph{accumulator}. The generator consists of a block cipher that
encrypts a counter value to produce a block of random bytes. After every data
request the encryption key is replaced with a new output from the block cipher
in order to reduce security vulnerabilities in the event an of attacker
learning of the internal state~\cite{fortuna}.

Beyond basic implementation details, \sysname includes a number of design
changes from the original specification of Fortuna: seed file management,
entropy pool inputs, and entropy pool size. 

\para{Seed File Management.} The originally specified manner for managing the
seed file is to rewrite the contents of the seed with an output from the
generator after every startup~\cite{fortuna}. On an embedded device this is
difficult due to the lack of an initial seed on the first boot of a device and
limited non-volatile memory. By using SRAM startup state (see
Section~\ref{ssec:sram}), we removed the need for managing a seed file.
Instead, \sysname reads the mixed SRAM state from the low 64 bytes of memory,
which is produced during a CRC startup routine before the main program begins.

This seed management scheme also improves the boot speed of the application
when compared to the alternative of writing and reading a seed file from FLASH.
FLASH operations take a considerable amount of time and would slow down the
device start up during seed file management. Using the CRC mixing function is
fast, as each calculation happens in hardware within 2 clock cycles while the
CPU prepares the next byte of SRAM for computation.

\para{Entropy Pool Contribution.} The accumulator consists of a set of 32
entropy pools which are fed random data from any number of entropy sources. In
Fortuna, the data contributed to each pool consists of the random data itself
and a source identifier which indicates the origins of the sample. However, as
memory is not readily available in embedded devices, \sysname does not include
the source IDs when adding the source sample to an entropy pool.

\para{Entropy Pool Size.} When a pool is filled with enough entropy to
replenish the generator, the pool contents are collapsed into a 256-bit value
that becomes the new generator key.  The collapsing is done using a
cryptographic hashing algorithm.   For \sysname, 58 events are required to
reseed the generator with 128 bits of entropy; this number was used as the
minimum entropy pool size for re-seeding in order to ensure consistent
protection of the generator state. \sysname's entropy sources can re-seed the
generator every \SI{3}{\ms}.

\subsection{Algorithm Selection}
\label{ssec:algorithms}

Fortuna's design does not specify a particular block cipher or hashing
algorithm, however there are limitations on which algorithms can be used due to
key, block, and digest sizes. The block cipher is required to have a 128-bit
block to fit a 128-bit counter value, and a 256-bit key. The hashing algorithm
digest must also be 256-bit in order to produce new keys during re-seeding.
Further limitations placed our \sysname variation  were code size and execution
time due to limited memory and a slow clock speed.

We selected \textbf{AES-256} and
\textbf{SHA-2 256}  as the block cipher and hashing algorithm,
respectfully, due to their input and output sizes, and the availability of
software libraries for embedded systems.

\begin{figure}[t]
	\centering
	\includegraphics[width=0.5\textwidth]{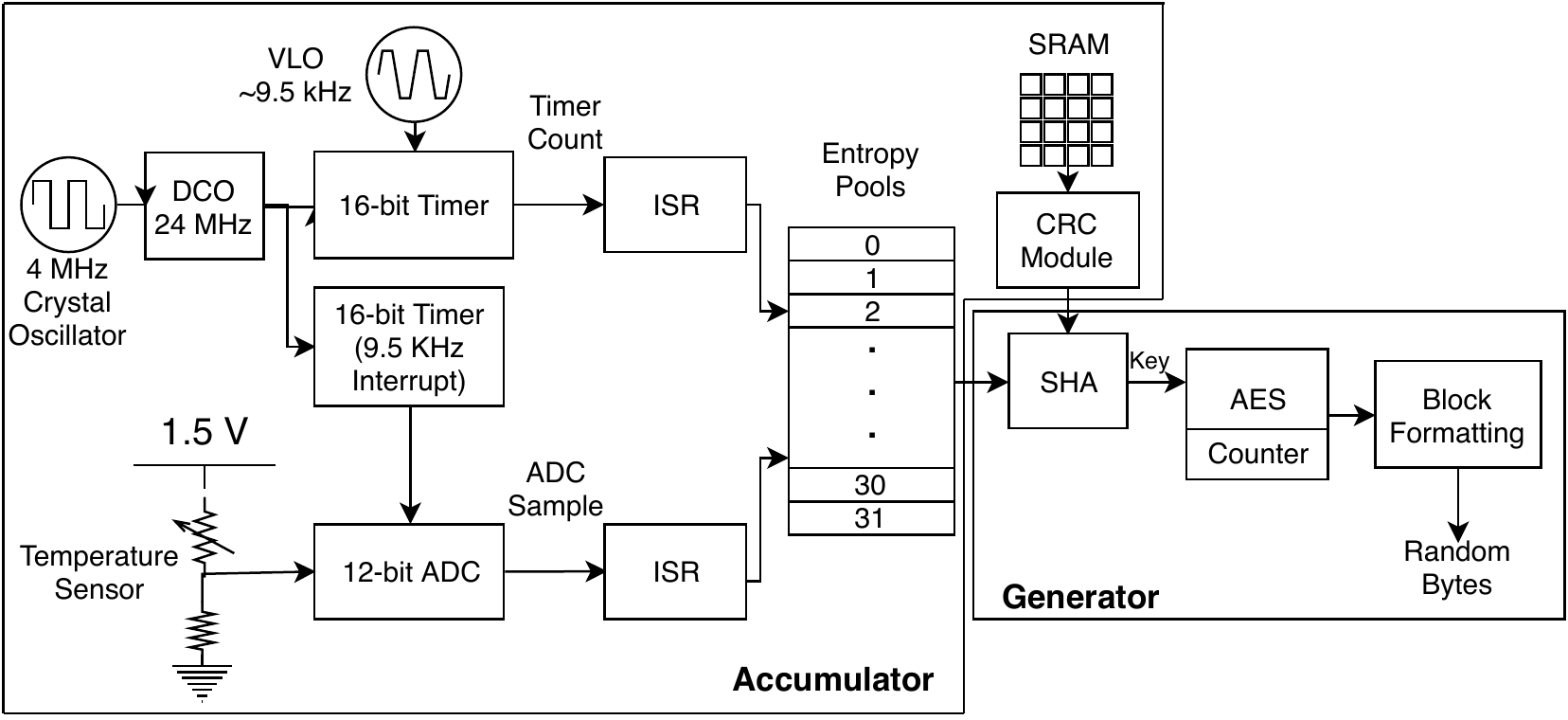}
	\caption{\textbf{\sysname Architecture.} }
	\label{fig:system}
\end{figure}

\subsection{Implementation}

The full \sysname architecture, as implemented on the MSP430 and illustrated in
Figure~\ref{fig:system}, utilizes two runtime entropy sources (the VLO and
temperature sensor) and one entropy source for the seed file (SRAM startup
state). The generator operates independently from the accumulator and
periodically hashes the entropy pools into the generator block cipher core.
The entropy sources are discussed in more detail in Section~\ref{sec:entropy}.

Taking into account the entropy estimates of both the VLO and the temperature
sensor, approximately 4.4 bits of entropy are collected between the two runtime
sources at an average rate of \SI{9.5}{\kHz}.  In the event that the internal
state of the generator is compromised, the accumulator would take 58 entropy
source events, or \SI{3}{\ms}, to gather 128 bits of entropy and reseed the
generator. 

We used the NIST Statistical Test Suite~\cite{nist-rand} to test the randomness
of \sysname and summarize the results in Table~\ref{tab:prng_rand}. Columns C1
to C10 represent the frequency at which passing p-values are calculated for
blocks (p-values gauge whether a block is likely to pass a specific randomness
threshold). The p-value calculated from a chi-square test is in the following
column, with the pass-rate of the test following in the next column. All of the
pass rates are above the required threshold for confirming good random data.

\begin{table*} \centering
	\caption{Randomness tests on \sysname samples}
  \sffamily
  \ra{1.3}
  \begin{tabular}{@{}lrrrrrrrrrrrc@{}}
    \toprule
    &\textbf{C1} & \textbf{C2} & \textbf{C3} & \textbf{C4} & \textbf{C5} &
		\textbf{C6} & \textbf{C7} & \textbf{C8} & \textbf{C9} & \textbf{C10} &
		\textbf{P-Value} & \textbf{Pass Ratio}\\
    \midrule
    \textbf{Frequency} & 288 & 285 & 262 & 265 & 296 & 268 & 239 & 327 & 266 & 304 & 0.015832 & 2771/2800  \\
		\textbf{BlockFrequency} &276 & 287 & 264 & 312 & 261 & 289 & 275 & 277 & 285 & 274 & 0.666097 & 2774/2800 \\
		\textbf{CUmulativeSums} &292 & 242 & 293 & 301 & 274 & 255 & 267 & 253 & 320 & 303 & 0.012556 & 2769/2800 \\
		\textbf{Runs} &260 & 285 & 323 & 273 & 276 & 267 & 261 & 287 & 269 & 299 & 0.205375 & 2774/2800 \\
		\textbf{LongestRun} &275 & 313 & 286 & 296 & 281 & 274 & 248 & 266 & 269 & 292 & 0.314759 & 2773/2800\\
    \bottomrule
	\end{tabular}
	\label{tab:prng_rand}
\end{table*}

 \section{Entropy Sources}
\label{sec:entropy}

\sysname leverages three components of the MSP430 as entropy sources for
seeding the generator and feeding the accumulator. Two of the sources, phase
jitter in an internal low-power oscillator and a temperature sensor were used
as feeders to the entropy pool. The third source, the startup state of SRAM,
was used as an initial seed for the generator. The following sections discuss
each of these sources and analyze their behavior and entropic qualities. 

\subsection{Low-Powered Oscillator}
\label{ssec:vlo}

The MSP430 has an internal Very Low-Power Oscillator (VLO) that is intended to
be used in low-power applications where an external oscillator is not able to
be powered or is not present. As it is powered internally and is not sourced
from a high-precision crystal, the VLO is subject to \emph{phase jitter}. Phase
jitter is the small time difference between when a controlled oscillator has a
rising edge, and when it is \textit{expected} to have a rising edge in an ideal
model~\cite{jitter}. This signal artifact has been used in other RNGs
before~\cite{osc-gen, osc-gen-cmos, sunar-martin}. 

One other characteristic of interest is oscillator wander, the tendency for the
frequency to stray far away from the typical frequency. Unlike phase jitter,
this phenomenon is not desirable in an entropy source, as oscillator wander can
be influenced by changes in the operating environment~\cite{jitter}.  Using an
oscilloscope capture in infinite-persist mode we observed that the period sits
at a consistent typical frequency with slight fluctuations in the form of phase
jitter, i.e., the VLO instability manifests as jitter and not oscillator
wander. 

\para{Entropy Estimate.} We used the NIST SP 800-90B entropy estimation suite
to measure a min-entropy of 1.47 bits per 8-bit sample at a voltage of
\SI{2.4}{\V} using one million samples. We also tested samples from the VLO
across a voltage range of \SI{2.4}{\V} to \SI{3.6}{\V}---the maximum supply
range for the MSP430 running at \SI{24}{\MHz}~\cite{msp430-ds}---and a
temperature range of \SI{75}{\degree F} to \SI{0}{\degree F}. The lowest
min-entropy we observed was 1.19 bits per sample at \SI{2.9}{\V} at
\SI{75}{\degree F}.

At a fixed temperature and voltage, the VLO min-entropy estimates remained
consistent, regardless of the temperature itself. However, when samples were
collected while the environment temperature was actively changing, the entropy
estimates decreased. This is likely because of the VLO periods increasing or
decreasing in a constant direction, introducing a more predictable pattern to
the period samples.  

\subsection{Temperature Sensor}
\label{ssec:temp}

The second runtime entropy source used by \sysname is an internal temperature
sensor that measures an aggregate of the environment and internal temperatures
of the microcontroller. The sensor was routed as an input to a 12-bit Analog to
Digital Converter (ADC) with a reference voltage of \SI{1.5}{\V} to increase
the resolution of the sensor to \SI{0.366}{\mV}/bit for maximum sensitivity. In
order to allow each entropy source to contribute to the pools evenly, we
configured the ADC to make conversions of the temperature sensor at
\SI{9.5}{\kHz} to match the typical period of the VLO at a supply of
\SI{3.3}{\V}. 

\para{Entropy Estimate.} The min-entropy estimate for the temperature sensor at
\SI{9.5}ksps was 2.93 bits per 8-bit sample.

\subsection{SRAM Startup State}
\label{ssec:sram}

In addition to the runtime  entropy sources, \sysname also requires a 64-byte
\textbf{seed file} that initializes the generator key at start up; this is a
challenge in embedded devices due to limited non-volatile memory. However,
previous work has shown that the startup state of SRAM cells exhibits a random
pattern and can be used as a large pool of boot-time entropy in embedded
systems~\cite{holcomb}. Our \sysname implementation utilized the start up state
of SRAM on the MSP430 (\SI{10}{KB}) o create a 64-byte seed file. Here we
define \emph{startup state} as the  initial memory state after power-on before
any function calls or data were placed on the software stack. 

\para{Entropy Estimate.} We measured the startup state across 100 startups with
30 second power-off periods between each startup to create a set of one million
samples. Although SRAM was estimated to have 0.457844 bits of entropy per byte,
we observed that bytes change at different rates; The majority of the bytes
changed from 3 to 10 times, while others changed up to 30 times.  This
observation suggest that the randomness is likely spread out over the entire
memory and it is important to use the entire memory  to calculate the
seed---i.e.,  \SI{10}{KB} needs to be collapsed down to a 64 byte memory
segment without disturbing the RAM state. To do this, \sysname uses a built-in
CRC module to mix each block of \SI{160}{bytes} down to a 16-bit value with
high entropy. 
The viability of this \textbf{CRC-CCITT-16} hardware mixing
function is discussed in the following section.

\begin{table*} \centering
    \caption{Linux source entropy estimates}
    \sffamily
  \begin{tabular}{@{}lrrrrr@{}}
    \toprule
     & \textbf{Min-Entropy} & \textbf{MCV Est.} & \textbf{Collision Est.} &
     \textbf{Markov Est.} & \textbf{Compression Est.}\\
    \midrule  
    \textbf{User Input ID} & 0.7419 & 2.0395 & 1.3447 & 1.8323 & 0.7419 \\
    \textbf{User Input Time} & 0.1147 & 7.8872 & 0.1147 & 1.0752 & 0.9996 \\
    \textbf{Interrupt ID} & 0.3817 & 0.9961 & 0.7572 & 0.9243 & 0.3817 \\
    \textbf{Interrupt Time} & 0.6000 & 5.3211 & 0.6692 & 0.6000 & 0.8555 \\
    \textbf{Disk Access ID} & 0.0075 & 0.9964 & 0.8398 & 0.0075 & 0.3845 \\
    \textbf{Disk Access Time} & 0.3159 & 5.2679 & 0.3159 & 0.4325 & 0.5134 \\
  \bottomrule
  \end{tabular}
  \label{tab:linux_entropy}
\end{table*}

\subsection{Comparison of Entropy Sources}
\label{ssec:entropy_comp}

As a performance and security benchmark, we compared the entropy sources in
\sysname  to those used by a  general purpose Linux system.  The Linux
generator uses three sources of entropy: User input events (keyboard and
mouse), hardware interrupts, and disk access events~\cite{lrng}.  Each event is
paired with a timestamp, the precision of which is dependent on the
system configuration; for our experiments  we observed  timestamps with
\SI{4}{\ms} precision. In addition to the entropy sources, the generator's
entropy pool is saved and restored across system power cycles to provide
starting entropy.

Using a modification to the Linux kernel, we collected one million samples from
each source to test with the NIST estimation suite.  Shown in
Table~\ref{tab:linux_entropy},  the maximum amount of entropy from a single
source was 0.742 bits per sample from user input events.  The time needed to
collect the high-entropy input events spanned three days, whereas interrupts
required 1 hour, and disk events required over a day for disk access events to
be initiated by the user.

The \sysname sources  have significantly higher estimates than the Linux
entropy sources, particularly the VLO and temperature sensor. Additionally,
both the user input and disk event sources are dependent on user input, which
could starve the system if the computer is not used frequently enough. In
contrast, the sources used in this work are not dependent on an external user.
Furthermore, the \sysname sources run much faster than the Linux sources, and
can generate a high volume of samples in a very short period of time (on the
order of seconds). 
\subsubsection{CRC as a Mixing Function}
\label{sssec:crc}

CRCs are a type of algebraic cyclic code that operates on the underlying
principles of polynomial division over the field GF(2), modulo a generator
polynomial $G(x)$. The individual bits of the CRC inputs are treated as the
binary coefficients of polynomials, and the remainder, often called the
syndrome, of division by $G(x)$ is used as the CRC output~\cite{crc-book}. This
structure of a cyclic code is important in the analysis of CRCs for entropy
mixing.

One class of entropy mixing functions is resilient functions. A [n, m, d]
resilient function maps from a domain of $Z_2^n$, to $Z_2^m$ (n-bit and m-bit
binary strings), where the knowledge or control over $d$ bits of the input
reveal no information about the output. Such functions can be formed from linear
error-correcting codes with Hamming Distance $d+1$. The mapping from domain to
codomain is performed by multiplying an input by a generator matrix, $G$, for
the linear code [$F(x) = xG^T$]. This operation is analagous to calculation of a
syndrome within a cyclic code~\cite{sunar-martin}, and makes the CRC-CCITT-16
an attractive candidate for a resilient function.

Since CRCs are error-correcting codes, they are designed to produce different
syndromes for different input sequences to allow for error detection. As a 16
bit CRC, this variation of the code is able to detect any bit errors (ie.
changing bits), within a 16-bit input~\cite{crc-book}. Thus, there is a unique
output for any inputs up to 16 bits, so individual oscillator period sample
inputs will have unique effects on the final CRC result. Furthermore, as
CRC-CCITT-16 is implemented as a Linear Feedback Shift Register (LFSR), the
nature of the output is randomized, as is the purpose of a LFSR-based
RNG~\cite{crypto-book}.

Beyond the results of individual inputs, longer input streams must also produce
different results even with few varying input bits. A generator
with a primitive polynomial factor of degree $n$ for a CRC guarantees detection
of 2-bit differences in input streams that are at most $2^n-1$ bits
apart~\cite{crc-book}. Thus it can be guaranteed that different CRC-CCITT-16
outputs will result for input streams that differ in 2 bits that are 32767
places away from each other, far larger than the maximum of 1280 bits that pass
through the CRC for extraction. Additionally, since the generator polynomial
$G(x) = x^{16}+x^{12}+x^5+1$ is the product of $(x+1)$ and a primitive
polynomial, the CRC can detect all parity errors, which are errors consisting of
an odd-number of bit differences~\cite{crc-book}.

Based on previous work in CRC analysis, the CCITT-16 generator with an input of
256 bits has a Hamming Distance of 4 bits; precisely 6587 inputs with 4-bit-wide
differences will be un-noticed~\cite{crc-polys}. Taking into account that a
large amount of the input bits from SRAM are deterministic and that the majority
of bytes only differ by a few values, a very small fraction of all the possible
input sequences may not be detected. Additionally, there are no 5-bit errors
that are undetectable~\cite{crc-polys}. Any 6-bit errors are highly unlikely to
occur due to limited number of values each SRAM byte may contain.

The relevant properties of CRC-CCITT-16 as a mixing function are summarized as
follows. \1 A primitive-based generator allows detection of all odd-numbered
bit differences. \2 Generator polynomial degree 15 guarantees detection of all
2-bit differences within the 256-bit input. \3 A maximal-length LFSR design
ensures pseudo-randomness in outputs.  \4 A small amount of 4-bit-wide input
differences are undetectable, although wide bit differences are not critical to
detect in this application.  \5 All 5-bit-wide input differences are
detectable.  \6 We don't expect any 6-bit-wide input differences. \section{Conclusion}
\label{sec:conclusion}

We presented the design of a PRNG for commonly available embedded systems,
\sysname, that uses existing hardware peripherals as entropy sources. A 
previously design PRNG variant, Fortuna, was used along with three entropy
sources: jitter in a low-power oscillator, measurements from an internal
temperature sensor, and the start up state of SRAM. We provided entropy
estimates for each source, and discussed how to build a system that efficiently
uses memory and processing time resources to manage the collection and
processing of sources. Furthermore, we proposed the use of a CRC hardware
module to offload processing of the SRAM startup state to obtain an initial seed
for the PRNG. Using \sysname as a reference, similar PRNGs can be implemented on
other platforms to achieve reliable randomness, enabling security-critical
functions in resource-restricted systems. 
\bibliographystyle{./IEEEtran}
\bibliography{../bib/bib}

\begin{thebibliography}{10}
\providecommand{\url}[1]{#1}
\csname url@samestyle\endcsname
\providecommand{\newblock}{\relax}
\providecommand{\bibinfo}[2]{#2}
\providecommand{\BIBentrySTDinterwordspacing}{\spaceskip=0pt\relax}
\providecommand{\BIBentryALTinterwordstretchfactor}{4}
\providecommand{\BIBentryALTinterwordspacing}{\spaceskip=\fontdimen2\font plus
\BIBentryALTinterwordstretchfactor\fontdimen3\font minus
  \fontdimen4\font\relax}
\providecommand{\BIBforeignlanguage}[2]{{%
\expandafter\ifx\csname l@#1\endcsname\relax
\typeout{** WARNING: IEEEtran.bst: No hyphenation pattern has been}%
\typeout{** loaded for the language `#1'. Using the pattern for}%
\typeout{** the default language instead.}%
\else
\language=\csname l@#1\endcsname
\fi
#2}}
\providecommand{\BIBdecl}{\relax}
\BIBdecl

\bibitem{fortuna}
N.~Ferguson and B.~Schneier, \emph{Practical Cryptography (2nd Edition)}.\hskip
  1em plus 0.5em minus 0.4em\relax New York, NY: John Wiley and Sons, 2003, pp.
  1--432.

\bibitem{osc-gen}
A.~Cherkaoui, V.~Fischer, L.~Fesquet, and A.~Aubert, ``A very high speed true
  random number generator with entropy assessment,'' in \emph{Cryptographic
  Hardware and Embedded Systems - CHES 2013}, G.~Bertoni and J.-S. Coron,
  Eds.\hskip 1em plus 0.5em minus 0.4em\relax Berlin, Heidelberg: Springer
  Berlin Heidelberg, 2013, pp. 179--196.

\bibitem{osc-gen-cmos}
S.~Robson, B.~Leung, and G.~Gong, ``Truly random number generator based on a
  ring oscillator utilizing last passage time,'' \emph{IEEE Transactions on
  Circuits and Systems II: Express Briefs}, vol.~61, no.~12, pp. 937--941, Dec
  2014.

\bibitem{holcomb}
D.~E. Holcomb, W.~P. Burleson, and K.~Fu, ``Power-up sram state as an
  identifying fingerprint and source of true random numbers,'' \emph{IEEE
  Transactions on Computers}, vol.~58, no.~9, pp. 1198--1210, Sept 2009.

\bibitem{nist-rand}
L.~E. Bassham, III, A.~L. Rukhin, J.~Soto, J.~R. Nechvatal, M.~E. Smid, E.~B.
  Barker, S.~D. Leigh, M.~Levenson, M.~Vangel, D.~L. Banks, N.~A. Heckert,
  J.~F. Dray, and S.~Vo, ``Sp 800-22 rev. 1a. a statistical test suite for
  random and pseudorandom number generators for cryptographic applications,''
  Gaithersburg, MD, United States, Tech. Rep., 2010.

\bibitem{jitter}
{Statek Corporation}, ``{An overview of oscillator jitter},''
  \url{http://statek.com/wp-content/uploads/2018/03/tn35-Rev-B.pdf}, 2007,
  online; accessed 20 February 2017.

\bibitem{sunar-martin}
\BIBentryALTinterwordspacing
B.~Sunar, W.~J. Martin, and D.~R. Stinson, ``A provably secure true random
  number generator with built-in tolerance to active attacks,'' \emph{IEEE
  Trans. Comput.}, vol.~56, no.~1, pp. 109--119, Jan. 2007. [Online].
  Available: \url{http://dx.doi.org/10.1109/TC.2007.4}
\BIBentrySTDinterwordspacing

\bibitem{msp430-ds}
\emph{MSP430F552x, MSP430F551x Mixed-Signal Microcontrollers datasheet}, Texas
  Instruments, 11 2015, rev. M.

\bibitem{lrng}
Z.~Gutterman, B.~Pinkas, and T.~Reinman, ``Analysis of the linux random number
  generator,'' in \emph{2006 IEEE Symposium on Security and Privacy (S P'06)},
  May 2006, pp. 15 pp.--385.

\bibitem{crc-book}
W.~H. Press, S.~A. Teukolsky, W.~T. Vetterling, and B.~P. Flannery, ``Cyclic
  redundancy and other checksums,'' in \emph{Numerical Recipces: The Art of
  Scientific Computing}, 3rd~ed.\hskip 1em plus 0.5em minus 0.4em\relax The
  Edinburgh Building, Cambridge CB2 8RU, UK: Cambridge University Press, 2007,
  ch. 22.4, pp. 1168--1175.

\bibitem{crypto-book}
C.~Paar and J.~Pelzl, ``Shift register-based stream ciphers,'' in
  \emph{Understanding Cryptography: A Textbook for Students and
  Practitioners}.\hskip 1em plus 0.5em minus 0.4em\relax Springer-Verlag Berlin
  Heidelberg, 2010, ch. 2.3, pp. 41--49.

\bibitem{crc-polys}
\BIBentryALTinterwordspacing
P.~Koopman, ``Crc polynomial zoo,''
  \url{http://users.ece.cmu.edu/~koopman/crc/crc16.html}, Nov 2016. [Online].
  Available: \url{http://users.ece.cmu.edu/~koopman/crc/crc16.html}
\BIBentrySTDinterwordspacing

\end{thebibliography}

\end{document}